\documentclass[10pt,twocolumn,a4paper]{article}

\usepackage{lmodern}
\usepackage[T1]{fontenc}
\usepackage[cp1250]{inputenc}
\usepackage{graphicx}
\usepackage{epstopdf}

\usepackage{varioref}
\usepackage{amsmath,amssymb,amsfonts}

\usepackage[pdftex,breaklinks=true]{hyperref}
\hypersetup{
colorlinks=true,
linkcolor=black,          
citecolor=black,        
filecolor=black,      
urlcolor=black,           
bookmarks=true,
bookmarksnumbered=true,
pdfauthor={Piotr Bania},
pdfstartview=FitH,
pdfsubject={JIT spraying and mitigations},
pdftitle={JIT spraying and mitigations}
}

\usepackage{breakurl}
\usepackage{url}
\usepackage{algorithmic}
\usepackage{listings}
\usepackage{color}
\usepackage[ruled,vlined]{algorithm2e}

\title{JIT spraying and mitigations}
\author{Piotr Bania\\Kryptos Logic Research\\
\texttt{\href{http://www.kryptoslogic.com}{www.kryptoslogic.com}}}
\date{2010}

\begin{document}
\maketitle

\begin{abstract}

With the discovery of new exploit techniques, novel protection mechanisms are needed as well. Mitigations like DEP (Data Execution Prevention) or ASLR (Address Space Layout Randomization) created a significantly more difficult environment for exploitation. Attackers, however, have recently researched new exploitation methods which are capable of bypassing the operating system's memory mitigations. One of the newest and most popular exploitation techniques to bypass both of the aforementioned security protections is JIT memory spraying, introduced by Dion Blazakis~\cite{DionJIT}.

In this article we will present a short overview of the JIT spraying technique and also novel mitigation methods against this innovative class of attacks. An anti-JIT spraying library was created as part of our shellcode execution prevention system.

\end{abstract}

\section{Introduction}

In order to increase the security level of the operating system, Microsoft has implemented several mitigation mechanisms including DEP and ASLR. Data Execution Prevention (DEP) is a security feature that prohibits the application from executing code from a non-executable memory area. To exploit a vulnerability, an attacker must first find a executable memory region and then be able to fill it with necessary data (i.e., shellcode instructions). Generally, achieving this goal using old exploitation techniques is made significantly harder with the addition of the DEP mitigation. As a result, attackers improved upon the classic return-into-libc technique and started using Return-Oriented Programming (ROP)~\cite{RopPresent,RopPaper} to bypass DEP. However, techniques like ROP still rely on the attacker understanding memory layout characteristics, leading Microsoft to implement ASLR as countermeasure. ASLR renders the layout of an application's address space less predictable because it relocates the base addresses of the executable modules and other memory mappings. The JIT spraying technique~\cite{DionJIT} was introduced to bypass ASLR and DEP simultaneously. In this article we present our novel mechanisms which are created specifically to prevent the JIT spraying technique from successful execution. This research targeted Microsoft Windows operating systems with the x86-32 CPU architecture. The mitigations specifically focus on the ActionScript JIT compiler, which is currently being heavily used for this type of attack.

\section{JIT Spraying}

There are two general reasons why JIT spraying is a very useful exploitation method. Firstly, the code generated by the JIT compiler is stored in memory marked as executable. This should be obvious because otherwise JIT compiler would be unable to work correctly on systems shipped with the DEP feature. Evidently, if the attacker's code is generated by JIT engine it will also reside in the executable area. In other words, DEP is not involved in the protection of code emitted by the JIT compiler. This is a very useful method since the memory was not marked as executable in prior approaches like normal heap spraying. The second reason JIT spraying is powerful is that attacker's code location can be predicted correctly \cite{DionJIT,AlexeyJIT}, so at this point ASLR is also no longer a big threat for the attacker. In this article we will focus specifically on detecting JIT code generation required for the address discovery methods discussed in the referenced citation. The reader is encouraged to have a full understanding of the referenced work.

\subsection{JIT Code Generation}\label{sec:jitgen}

Just-In-Time compilation converts code at runtime; typically from bytecode into machine code. By doing this, an interpreted program's performance greatly improves. The JIT spraying method ``forces'' the JIT compiler to produce a lot of executable pages with embedded attacker's code.  In order to write the code to specific location, the JIT compiler must first mark the destination memory as writable. Since multiple generated code chunks may reside on the same memory page, the JIT compiler marks the entire page as RWX (Readable-Writable-Executable). These permissions are necessary because a different chunk of memory residing on the same page may be executed asynchronously (for example by a different thread), resulting in access violation if the requested memory page was not executable at that moment. After the code is written the compiler marks the destination region as RX - readable and executable not writable anymore, as shown in Listing~\ref{antiread}.

\begin{figure}
{\ttfamily{\footnotesize{
\lstset{language={[x86masm]Assembler}}
\begin{lstlisting}[frame=trbl, label=antiread, caption={Sample listing of changes of memory rights requested by JIT compiler.}, captionpos=b]{}
Requests:
MEM: 0x057d0090 size=1 prot=RWX
MEM: 0x057d0090 size=c prot=RX

Generated code:
0x057d0090 mov edx,[esp+0ch]
0x057d0094 mov ecx,[edx]
0x057d0096 call 0fdea9d1ah
0x057d009b ret

Requests:
MEM: 0x057d0170 size=1  prot=RWX
MEM: 0x057d0170 size=1a prot=RX

Generated code:
0x057d0170 mov edx,[esp+0ch]
0x057d0174 push dword [edx+0ch]
0x057d0177 push dword [edx+08h]
0x057d017a push dword [edx+04h]
0x057d017d mov ecx,[edx]
0x057d017f call 0fdea5601h
0x057d0184 mov eax,04h
0x057d0189 ret

\end{lstlisting}
}}}
\end{figure}

In order to force the JIT compiler to generate code that includes shellcode data, attackers must make use of ActionScript operators. Even though ActionScript consists of multiple operators like: arithmetic, arithmetic compound assignment, bitwise etc. only one appears to be used in the currently known shellcodes. Listing~\ref{jit_code_gen} presents generated code for few different types of operators (expression used: {\tt{a OP b OP c OP d ...}}). As a test-case, the data values we have used come from one of the very few available JIT shellcodes~\cite{AlexeyJIT}.

\begin{figure}
{\ttfamily{\footnotesize{
\lstset{language={[x86masm]Assembler}}
\begin{lstlisting}[frame=trbl, label=jit_code_gen, caption={Code generated by JIT compiler depending on the used operator.}, captionpos=b]{}
Operator XOR (^):
[b8 90 90 90 3c ] mov eax,03c909090h
[35 90 90 90 3c ] xor eax,03c909090h
[35 90 90 90 3c ] xor eax,03c909090h
[35 90 90 90 3c ] xor eax,03c909090h
[35 90 90 90 3c ] xor eax,03c909090h
[35 90 90 90 3c ] xor eax,03c909090h
[35 90 90 90 3c ] xor eax,03c909090h
...entire block of xors...
[35 31 d2 58 3c ] xor eax,03c58d231h
[35 80 ca ff 3c ] xor eax,03cffca80h
...

Operator ADD (+):
[b8 90 90 90 3c ] mov eax,03c909090h
[f2 0f 2a c0 ] cvtsi2sd xmm0,eax
[66 0f 28 c8 ] movapd xmm1,xmm0
[f2 0f 58 c8 ] addsd xmm1,xmm0
[f2 0f 58 c8 ] addsd xmm1,xmm0
...addsd...
[b8 31 d2 58 3c ] mov eax,03c58d231h
[f2 0f 2a c0 ] cvtsi2sd xmm0,eax
[f2 0f 58 c8 ] addsd xmm1,xmm0
[b8 80 ca ff 3c ] mov eax,03cffca80h
[f2 0f 2a c0 ] cvtsi2sd xmm0,eax
[f2 0f 58 c8 ] addsd xmm1,xmm0
...so on...

Operator MUL (*):
[b8 90 90 90 3c ] mov eax,03c909090h
[f2 0f 2a c0 ] cvtsi2sd xmm0,eax
[66 0f 28 c8 ] movapd xmm1,xmm0
[f2 0f 59 c8 ] mulsd xmm1,xmm0
[f2 0f 59 c8 ] mulsd xmm1,xmm0
...mulsd...
[b8 31 d2 58 3c ] mov eax,03c58d231h
[f2 0f 2a c0 ] cvtsi2sd xmm0,eax
[f2 0f 59 c8 ] mulsd xmm1,xmm0
[b8 80 ca ff 3c ] mov eax,03cffca80h
[f2 0f 2a c0 ] cvtsi2sd xmm0,eax
[f2 0f 59 c8 ] mulsd xmm1,xmm0
...so on...

Operator DIV (/):
[b8 90 90 90 3c ] mov eax,03c909090h
[f2 0f 2a c0 ] cvtsi2sd xmm0,eax
[66 0f 28 c8 ] movapd xmm1,xmm0
[f2 0f 5e c8 ] divsd xmm1,xmm0
[f2 0f 5e c8 ] divsd xmm1,xmm0

...divsd...
[b8 31 d2 58 3c ] mov eax,03c58d231h
[f2 0f 2a c0 ] cvtsi2sd xmm0,eax
[f2 0f 5e c8 ] divsd xmm1,xmm0
[b8 80 ca ff 3c ] mov eax,03cffca80h
[f2 0f 2a c0 ] cvtsi2sd xmm0,eax
[f2 0f 5e c8 ] divsd xmm1,xmm0
...so on...

\end{lstlisting}
}}}
\end{figure}

Listing~\ref{jit_code_gen} shows that when it comes to ActionScript operators, only {\tt{XOR}} appears to produce desirable results. For example, with the {\tt{XOR}} operator, the attacker controls four bytes of every single instruction. In other cases the expression arguments do not provide precise and predictable control over the emitted code blocks. By supplying different arguments to the expression it is possible to change the contents of specified blocks and make them more dependable on attacker's arguments, but the {\tt{XOR}} operator appears to be the best option for shellcode usage and that is probably why every known JIT shellcode makes use of this operator. Once the attacker is able to spray controlled executable instructions into the heap, the rest of the exploitation process goes the standard route. The main idea here is to spray the memory with instructions that include attacker's payload and then be able to transfer the execution there (like for example be able to point the instruction pointer ({\tt{EIP}}) to the address of {\tt{xor eax, IMM32}} operand).

\section{Mitigations}\label{mitigations}

In order to stop JIT spraying attacks we must be able to decide whether the code generated by the Just-In-Time engine should be marked as shellcode or not. This is not a trivial task, since the shellcode detector must not heavily impact the original program performance and it must also be free of false positive alerts. At this point there are two general approaches to implement in the shellcode detector:

\begin{itemize}
    \item Signature detection approach (scanning for NOP sleds, GetPC code, decoding chunk (decryption) codes etc.)
    \item Heuristic detection approach.
\end{itemize}

Signature based detectors are the most simple to implement but they also tend to generate a high number of false positive alerts. Signature detection is often not enough since the attacker may able to bypass it by constructing the code in another fashion~\cite{NIDS_poly_evasion, CLET,Patton01anachilles,Bania:1183364}. To make the detection process more reliable and less static we have decided to use the heuristic detection approach.

As shown in~\autoref{sec:jitgen}, before the JIT generated code is executed the memory protections need to be changed to RX (Read-Executable). To achieve this, the JIT engine executes the {\tt{VirtualProtect}} API function. It appears that the JIT generated code always starts at the memory address specified as a parameter to the {\tt{VirtualProtect}} API. This can be a serious advantage for detection purposes since at this point we know the selected address (API argument) is always a valid starting point (entry point) for disassembling. A second very useful factor here is that the generated code that typically contains embedded shellcode is generated in a very special fashion (see~\autoref{jit_code_gen}):\newline

{{\ttfamily{\footnotesize{
\lstset{language={[x86masm]Assembler}}
\begin{lstlisting}[frame=trbl, label=jit_c_scheme, caption={General structure of JIT generated code used in JIT spraying.}, captionpos=b]{}
mov 		reg,IMM32
operation	reg,IMM32

So in case of XOR operator:
mov		eax,IMM32
xor		eax,IMM32
xor		eax,IMM32
...

\end{lstlisting}
}}}

Our heuristic detection method is simple but very reliable. As shown in~\autoref{jit_c_scheme}, every JIT shellcode currently available generates instructions that use 32 bit immediate values as a source operand and the destination operand of such instruction is always a register that was previously initialized with another 32 bit immediate value. The initialization instruction is typically a {\tt{mov reg,IMM32}} or in most of the cases a {\tt{mov eax,IMM32}}. Our detection algorithm is described as follows (Algorithm \ref{algo_antijit}):

\begin{algorithm}
\label{algo_antijit}
\dontprintsemicolon
\SetKwInOut{Input}{input}
\SetKwInOut{Output}{output}
\Input{$region\_addr$, $region\_size$}
\Output{detection marker}
\Begin{
\ForEach{found\_mov\_imm32}{
$num_{instr} \longleftarrow 0$\;
$num_{badinstr} \longleftarrow 0$\;

\While{$num_{instr}$ < $MAX_{inum}$}{\;
$instr \longleftarrow$ disasm\_next\_instr()\;
\If{!instr or is\_terminator(instr)}{
break\;
}
\If{uses\_imm32\_operands(instr)}
{
$num_{badinstr} \longleftarrow num_{badinstr} + 1$\;
}
$num_{instr} \longleftarrow num_{instr} + 1$\;
}
\If{$num_{badinstr} > MAX_{ibadnum}$}
{
report\_shellcode()\;
}
}
}
\caption{JIT shellcode detection}
\end{algorithm}

where:
\begin{itemize}
    \item $num_{instr}$ - represents the number of disassembled instructions
    \item $num_{badinstr}$ - represents the number of "bad" instructions, in other words instructions that use 32 bit immediate operands
    \item $is\_terminator$ - represents a function which checks if the currently disassembled instruction should be marked as terminator (instructions like {\tt{CALL}},{\tt{RET}},{\tt{JMP}} and so on should be considered as terminators
    \item $uses\_imm32\_operands$ - represents a function which checks if currently processed instruction uses 32 bit immediate operands as source
    \item $MAX_{ibadnum}$ - represents a static number which describes the maximum number of "bad" instructions in the disassembled block
\end{itemize}

Our algorithm calculates the number of bad instructions (instructions that use 32 bit immediate operands as source) starting from the initialization instruction. This algorithm does not assume any specific destination register (so no matter if {\tt{EAX}} is used or any other x86-32 CPU register). Additionally it keeps counting the number of bad instructions even if they are separated by some other instruction(s) that does not use 32 bit immediate operands (for example by some MMX/SIMD instruction or any other which does not use 32 bit immediate operand as long as it is not a block terminator). It is also worth adding that entire scanning procedure takes place before attacker will have a chance to use the code generated by JIT engine since we constantly monitor all of the newly generated regions.

One might argue it would be easier to start the disassembly from the entry point instead of searching for {\tt{mov reg,IMM32}}. This is a more expensive approach, however, since the speed of that method would depend directly on the size of the block generated by JIT engine --- the longer the block, the slower the algorithm. Secondly, disassembling is always a very costly process when it comes to performance. By searching for {\tt{mov reg,IMM32}} and starting from that point we do not have to perform the entire region's disassembly.

\section{Countermeasures}

As we have described before in \autoref{sec:jitgen}, the attacker typically controls the 32 bit immediate operands. This gives him an opportunity to use 24 bits of the controlled value to encode shellcode instructions (since one byte is typically already used to perform a semantic NOP (i.e., {\tt{CMP AL}})). In order to bypass our protection, an attacker could try to produce his shellcode by creating multiple lands and linking them with short {\tt{JMP/JCC}} jumps (since they are only 2 bytes long). By doing this, an attacker can reduce the number of emitted instructions that use 32 bit immediate operands and relocate them through different memory regions. However, we have already created a working mitigation for this technique by additional scanning of the immediate value for {\tt{JMP/JCC}} opcodes. Since those jumps are always short attacker can only jump into a location between -128 to +127 bytes from the jump instruction. Considering the fact that by using this method attacker loses additional space for valuable shellcode instructions, he must emit a relatively large number of jump instructions. In our method we are not only scanning for {\tt{JMP/JCC}} opcodes inside of the immediate value, but we also check if the destination address of such jump points to a valid location that also includes another jump instruction and so on. In other words, we are trying to validate the potential jump chain. This step is necessary in order to avoid false-positive alerts. By connecting this approach with the previously described one in \autoref{mitigations} we have made the JIT spraying mitigation a lot harder to bypass.

\section{Testimonials}

Tests showed that anti JIT spraying protection library have not generated any false-positive alerts when browsing typical sites overloaded with ActionScript and flash animations. The library itself has not produced any noticeable changes to the original application's performance. All tests were performed on Flash version 10c (on Microsoft Windows Vista and XP operating systems).

\section{Conclusion}

In this article we have presented basic concepts of JIT memory spraying. This technique is a good countermeasure against protection mechanisms like DEP and ASLR and moreover it is getting more and more popular. In order to stop JIT spraying attacks from successful exploitation we have created mitigation techniques that are solid, reliable and fast. We hope the reader found this article interesting.

\section*{Acknowledgments}

The author would like to thank Richard Johnson, Dion Blazakis and Kryptos Logic team for reviewing this article.

We want to dedicate this article to the memory of Seba Jun aka "Nujabes" (1974 --- 2010).

\bibliographystyle{plain}
\bibliography{bibliografia}
\end{document}